\documentclass[12pt,a4paper]{article}
\usepackage{fullpage}
\usepackage{amsbsy}
\usepackage{amsfonts}
\usepackage{amssymb}
\begin{document}
\begin{titlepage}
\begin{flushright}
\begin{tabular}{l}
hep-ph/0104148
\\
15 April 2001
\end{tabular}
\end{flushright}
\vspace{1cm}
\begin{center}
\large\bfseries
Energy and Momentum of Oscillating Neutrinos
\\[0.5cm]
\normalsize\normalfont
C. Giunti
\\
\small\itshape
INFN, Sezione di Torino,
\\
\small\itshape
and
\\
\small\itshape
Dipartimento di Fisica Teorica,
Universit\`a di Torino,
\\
\small\itshape
Via P. Giuria 1, I--10125 Torino, Italy
\end{center}
\begin{abstract}
It is shown that Lorentz invariance implies that in general
flavor neutrinos in oscillation experiments
are superpositions of massive neutrinos with different energies
and different momenta.
It is also shown that for each process in which neutrinos are produced
there is either a Lorentz frame
in which all massive neutrinos have the same energy
or a Lorentz frame in which
all massive neutrinos have the same momentum.
In the case of neutrinos
produced in two-body decay processes,
there is a Lorentz frame
in which all massive neutrinos have the same energy.
\end{abstract}
\end{titlepage}

Neutrino oscillations is one of the most interesting
phenomena under investigation
in high-energy physics.
It gives information on neutrino masses and mixing,
that are fundamental ingredients for the
understanding of the Standard Model and its possible extensions.

The theory of neutrino oscillations has been studied
by many authors,
starting from Pontecorvo's pioneering works
\cite{Pontecorvo-mesonium-57,%
Pontecorvo-58,%
Pontecorvo-68,%
Gribov-Pontecorvo-69},
going through the classical works of the 70's
\cite{Eliezer-Swift-MIXING-76,%
Fritzsch-Minkowski-OSCILLATIONS-76,%
Bilenky-Pontecorvo-AGAIN-76,%
Bilenky-Pontecorvo-PR-78}
and the introduction of the wave packet description
\cite{Nussinov-coherence-76,
Kayser-oscillations-81,%
Giunti-Kim-Lee-Whendo-91,%
Giunti-Kim-Lee-Lee-93},
continuing until today with several new developments
(see \cite{Giunti-Kim-QMNO-00} and references therein).

In neutrino oscillation experiments
flavor neutrinos,
produced and detected in weak interaction processes,
are described by the flavor states
$|\nu_{\alpha}\rangle$,
with
$\alpha = e, \mu, \tau$,
which are superpositions of the states
of neutrinos with definite mass
(see \cite{Giunti-Kim-Lee-Remarks-92,%
CWKim-book,%
BGG-review-98,%
Bilenky-Giunti-lep-01}
and references therein):
\begin{equation}
|\nu_{\alpha}\rangle
=
\sum_k U_{\alpha k}^{*} \, |\nu_{k}\rangle
\,,
\label{01}
\end{equation}
where $U$ is the mixing matrix of the neutrino fields
($\nu_{\alpha} = \sum_k U_{\alpha k} \, \nu_{k}$)
and the index $k=1,2,\ldots$ labels the neutrino with mass $m_k$.
Neutrino oscillations are due to the
different phase velocities of different massive neutrinos,
leading to the transition probability
\begin{equation}
P_{\nu_\alpha\to\nu_\beta}(L,T)
=
\left|
\sum_k
U_{\alpha k}^{*}
\,
e^{i p_k L - i E_k T}
\,
U_{\beta k}
\right|^2
\,,
\label{02}
\end{equation}
where
$p_k$ and $E_k$ are the momentum and energy of the neutrino with mass $m_k$
and
$L$ and $T$ are the space and time intervals
between neutrino production and detection.
An important characteristic of the transition probability (\ref{02})
is its manifest Lorentz invariance.
Indeed, different observers must measure the same
flavor transition probability.

One of the problems under debate is
the determination of the mass dependence of the momenta and energies,
$p_k$, $E_k$,
of massive neutrinos in the transition probability (\ref{02}).
In the classical works of the 70's
\cite{Eliezer-Swift-MIXING-76,%
Fritzsch-Minkowski-OSCILLATIONS-76,%
Bilenky-Pontecorvo-AGAIN-76,%
Bilenky-Pontecorvo-PR-78}
it was assumed that the massive neutrinos
have a common momentum,
$p_k=p$,
and different energies given by
\begin{equation}
E_k = \sqrt{p^2+m_k^2} \simeq
p + \frac{m_k^2}{2p}
\,.
\label{001}
\end{equation}
The approximation is due to the extreme relativistic character
of detectable neutrinos, which implies also that $T \simeq L$,
leading to the standard expression
\begin{equation}
P_{\nu_\alpha\to\nu_\beta}(L)
=
\sum_k
|U_{\alpha k}|^2
|U_{\beta k}|^2
+
2
\mathrm{Re}
\sum_{k>j}
U_{\alpha k}^*
U_{\beta k}
U_{\alpha j}
U_{\beta j}^*
\,
\exp\left[
- i \frac{\Delta{m}^2_{kj}}{2p} \, L
\right]
\,,
\label{017}
\end{equation}
for the probability of
$\nu_\alpha\to\nu_\beta$ transitions
as a function of the source-detector distance $L$ measured in real experiments.
Here
$\Delta{m}^2_{kj} \equiv m_k^2 - m_j^2$.

As discussed in \cite{Giunti-Kim-QMNO-00}
and references therein,
formally it is possible to derive the transition probability
(\ref{017})
also assuming
that the massive neutrinos
have a common energy.
This assumption has the advantage to avoid
the approximation $T \simeq L$
in the derivation of the transition probability
(\ref{017}) as a function of the distance $L$
from the probability (\ref{02}),
which depends also on the time interval $T$
that is not measured in oscillation experiments.

However,
it is well known
\cite{Winter-81,%
Giunti-Kim-Lee-Whendo-91,%
Giunti-Kim-QMNO-00,
Bilenky-Giunti-lep-01}
that the assumption of equal momentum or equal energy
is not necessary for the derivation of the neutrino
oscillation probability (\ref{017}),
and actually it may be incompatible with energy-momentum
conservation in the process in which neutrinos are produced.
In general energy-momentum
conservation in the production process
implies that different massive neutrinos
have different momenta, $p_k$,
as well as different energies, $E_k$,
related by the relativistic dispersion relation
\begin{equation}
E_k^2 = p_k^2 + m_k^2
\,.
\label{002}
\end{equation}

Nevertheless,
some authors
\cite{Grossman-Lipkin-spatial-97,%
Lipkin-two-slit-00,%
Stodolsky-unnecessary-98}
claim that there is one correct assumption:
equal energy.

Here we present a simple argument that shows that
the equal energy assumption,
as well as the equal momentum assumption,
in general do not correspond to reality:
Lorentz invariance implies that even
if different massive neutrinos have the same energy
(momentum) in one Lorentz frame,
they have different energies (momenta)
in all the other frames boosted along the
neutrino propagation path.

Indeed,
let us assume for example that in a Lorentz frame $S$
different massive neutrinos have the same energy:
\begin{equation}
E_k = E
\,,
\label{003}
\end{equation}
independent from the mass index $k$.
In this frame the momenta of the massive neutrinos are given by
\begin{equation}
p_k = \sqrt{ E^2 - m_k^2 }
\simeq
E - \frac{m_k^2}{2E}
\,.
\label{004}
\end{equation}
Therefore $E$ is the momentum of a massless neutrino,
equal to its energy.
Since in oscillation experiments neutrinos
propagate along a macroscopic distance
between production and detection,
we will consider in the following only one spatial direction
along the neutrino path.

In another Lorentz frame $S'$ with velocity $v$
with respect to $S$ along the neutrino path
the energy of the $k^{\mathrm{th}}$ massive neutrino is
\begin{equation}
E'_k
=
\sqrt{\frac{1+v}{1-v}} \, E
-
\frac{v}{\sqrt{1-v^2}} \, \frac{m_k^2}{2E}
=
E' - \frac{v}{1-v} \, \frac{m_k^2}{2E'}
\,,
\label{005}
\end{equation}
where
\begin{equation}
E'
=
\sqrt{\frac{1+v}{1-v}} \, E
\label{006}
\end{equation}
is the energy of a massless neutrino in $S'$.
The difference between the energies of the
$k^{\mathrm{th}}$ and $j^{\mathrm{th}}$ massive neutrinos
is
\begin{equation}
\Delta E'_{kj} \equiv
E'_k - E'_j
= - \frac{v}{1-v} \, \frac{\Delta{m}^2_{kj}}{2E'}
\,.
\label{007}
\end{equation}
For relativistic velocities
($v \sim 0.1 - 1$),
the energy difference is of the same order as the momentum
difference,
\begin{equation}
\Delta p'_{kj} \equiv
p'_k - p'_j
= - \frac{1}{1-v} \, \frac{\Delta{m}^2_{kj}}{2E'}
\,.
\label{008}
\end{equation}
Therefore,
it is clear that in the Lorentz frame $S'$
the energies of different massive neutrinos are different
and the equal energy assumption is untenable.

For example,
let us consider the simple case of pion decay,
\begin{equation}
\pi^+ \to \mu^+ + \nu_\mu
\,.
\label{009}
\end{equation}
For the sake of illustration,
let us consider the equal energy assumption to be valid
for pion decay at rest,
even if this assumption is incompatible with
energy-momentum conservation
\cite{Winter-81}.
Then $S$ is the Lorentz frame in which the pion is at rest.

Many experiments measure the oscillations
of neutrinos produced by pion decay in flight.
These are short and long baseline accelerator experiments
and atmospheric neutrino experiments
(see \cite{BGG-review-98} for a review).
The energy of the pions goes from a few hundred MeV
(for example in the short baseline accelerator experiment
LSND \cite{LSND-flight-98})
to hundreds of GeV
(for example in the upward-going
events measured in the Super-Kamiokande atmospheric neutrino experiment
\cite{SK-upmu-99}).

It is clear that even if the equal energy assumption is valid
for pion decay at rest,
it cannot be valid even approximately
in the case of short and long baseline accelerator experiments
and atmospheric neutrino experiments.
Indeed,
considering for example
a neutrino emitted in the forward direction
by a pion decaying in flight with energy
$E_{\pi} \simeq 200 \, \mathrm{MeV}$,
the laboratory frame $S'$ is boosted with respect to the frame $S$
in which the pion is at rest by a velocity
$v \simeq 0.71$,
which gives
\begin{equation}
\frac{v}{1-v} \simeq 2.4
\,,
\qquad
\frac{1}{1-v} \simeq 3.4
\,.
\label{010}
\end{equation}
From Eqs.(\ref{007}) and (\ref{008})
one can see that the energy and momentum difference
between different massive neutrinos is of the same order of magnitude.
Obviously,
increasing the pion energy,
the energy and momentum differences increase and
tend to the same limit.

Let us emphasize that one would obtain the same result
choosing another Lorentz frame
in which the energies of different massive neutrinos
are assumed to be equal:
from Lorentz invariance
the equal energy assumption
cannot be simultaneously valid for all neutrino oscillation
experiments in which neutrinos are produced by pion decay
and it cannot be even valid in one experiment in which
the decaying pion have a spectrum of energies
(as always happens in practice).

Another obvious problem of the equal energy assumption,
as well as the equal momentum assumption,
is the arbitrariness of the choice of the Lorentz frame
in which it is valid,
which is not based on any physical argument.

Let us discuss now the effect of energy-momentum
conservation in the production process
on the energies and momenta of different massive neutrinos.
In the wave-packet description
of neutrino oscillations
energy-momentum
conservation in the production process
is compatible with the localization in space and time
of the production process
\cite{Giunti-Kim-Lee-Whendo-91,%
Giunti-Kim-Lee-Lee-93,%
Giunti-Kim-Lee-Whendo-98,%
Giunti-Kim-Coherence-98}.
For example,
in pion decay
the pion is described by a localized wave packet.
The average energy and momentum of the pion wave packet
determines the average energies and momenta
of the different massive neutrinos through energy-momentum conservation
and the size of the pion wave packet
determines the sizes of the neutrino wave packets.

Since all detectable neutrinos are extremely relativistic,
only the first order approximation in the mass contribution
to the energies and momenta
of the different massive neutrinos is relevant.
At zeroth order in the mass contribution
all neutrino masses are considered negligible
and all energies and momenta of massive neutrinos
in a Lorentz frame $S$ are equal:
$p_k = E_k = E$.
The value of $E$ is determined by energy-momentum conservation
in the production process.
For example,
in pion decay at rest
$
E
=
\frac{ m_{\pi} }{ 2 }
\left( 1 - \frac{ m_{\mu}^2 }{ m_{\pi}^2 } \right)
\simeq
30 \, \mathrm{MeV}
$.

Since the energy $E_k$ and momentum $p_k$
of the $k^{\mathrm{th}}$
massive neutrino
are related by the relativistic dispersion relation (\ref{002}),
the first order corrections to the equalities
$p_k = E_k = E$
depend on the square of the neutrino mass.
In general,
energy-momentum
conservation in a Lorentz frame $S$
implies that
\begin{equation}
p_k
\simeq
E
-
\xi
\,
\frac{ m_{k}^2 }{ 2 E }
\,,
\label{0061}
\end{equation}
where $\xi$ is a quantity that depends on the production process.
For example,
in pion decay at rest
$
\xi
=
\frac{1}{2}
\left( 1 + \frac{m_\mu^2}{m_\pi^2} \right)
\simeq
0.8
$.
From the relativistic approximation
of the energy-momentum dispersion relation (\ref{002}),
the energy $E_k$ of the $k^{\mathrm{th}}$ massive neutrino
is given by
\begin{equation}
E_k
\simeq
E
+
\left( 1 - \xi \right)
\frac{ m_{k}^2 }{ 2 E }
\,.
\label{0062}
\end{equation}
From Eqs.(\ref{0061}) and (\ref{0062})
it is clear that the equal momentum and equal energy assumptions
correspond, respectively,
to the special cases $\xi=0$ and $\xi=1$.
However,
it is important to remember that $\xi$
is determined by energy-momentum conservation
in the production process and
in general its value is different from 0 or 1,
as we have seen in the case of pion decay at rest.

Let us consider now a Lorentz frame $S'$
boosted by a velocity $v$ along the neutrino propagation path.
The energies and momenta of the massive neutrinos
in the frame $S'$ can be written as
\begin{equation}
E'_k
\simeq
E'
+
\left( 1 - \xi' \right)
\frac{ m_{k}^2 }{ 2 E' }
\,,
\qquad
p'_k
\simeq
E'
-
\xi'
\,
\frac{ m_{k}^2 }{ 2 E' }
\,,
\label{011}
\end{equation}
with $E'$ given by Eq.~(\ref{006}) and
\begin{equation}
\xi'
=
\frac{\left(1+v\right) \xi - v}{1-v}
\,.
\label{012}
\end{equation}

In general
different massive neutrinos have different energies and momenta
also in the Lorentz frame $S'$,
but it may be possible to find a frame
in which the equal energy assumption corresponds to reality,
\begin{equation}
\xi' = 1
\quad
\Longleftrightarrow
\quad
v
=
\frac{1-\xi}{\xi}
\,,
\label{013}
\end{equation}
or another frame in which the equal momentum assumption
corresponds to reality,
\begin{equation}
\xi' = 0
\quad
\Longleftrightarrow
\quad
v
=
\frac{\xi}{1-\xi}
\,,
\label{014}
\end{equation}

Since $|v|<1$,
for a given process, which determines the value of $\xi$,
only one of Eqs.~(\ref{013}) or (\ref{014})
can be satisfied.
A frame in which
the equal energy assumption corresponds to reality exists if
$\xi > 1/2$,
whereas
a frame in which
the equal momentum assumption corresponds to reality exists if
$\xi < 1/2$.
From Eq.~(\ref{012}) one can see that, consistently,
if
$\xi > 1/2$,
also
$\xi' > 1/2$ in any frame
and
if
$\xi < 1/2$,
also
$\xi' < 1/2$ in any frame
($\xi = 1/2$ implies
$\xi' = 1/2$ in any frame).

It is remarkable that
if there is a Lorentz frame in which
the condition (\ref{013})
is satisfied,
in this frame
all massive neutrinos have the same energy,
whatever their number.
Similarly,
if there is a Lorentz frame in which
the condition (\ref{014})
is satisfied,
in this frame
all massive neutrinos have the same momentum.
This is due to the first order relativistic approximations
(\ref{0061}) and (\ref{0062})
and to the linearity of Lorentz transformations,
that imply a similar expression for $p_k$ and $E_k$
in any frame (confront Eq.~(\ref{011})
with Eqs.~(\ref{0061}) and (\ref{0062})).

In the case of pion decay at rest,
we have seen that $\xi \simeq 0.8$.
Therefore there is no frame in which
the equal momentum assumption corresponds to reality,
whereas for neutrinos produced in the forward direction
in the decay of pions with velocity $v \simeq 0.25$ and energy
$E'_{\pi} \simeq 145 \, \mathrm{MeV}$,
the equal energy assumption corresponds to reality.

The non-existence of a frame in which
the equal momentum assumption corresponds to reality
and the existence of a frame in which
the equal energy assumption corresponds to reality
is a property of all two body decay processes in which neutrinos
are produced.
Indeed in the general two body decay process
\begin{equation}
A^+ \to \alpha^+ + \nu_\alpha
\qquad
(\alpha=e,\mu,\tau)
\label{015}
\end{equation}
at rest, $\xi$ is given by
\begin{equation}
\xi
=
\frac{1}{2}
\left( 1 + \frac{m_\alpha^2}{m_A^2} \right)
>
\frac{1}{2}
\,,
\label{016}
\end{equation}
where $m_A$ and $m_\alpha$ are, respectively,
the masses of the decaying particle $A$ and of the charged lepton $\alpha$.
All massive neutrinos have
equal energy in the Lorentz frame where
the initial particle $A$ has velocity and energy given by
\begin{equation}
v
=
\frac{m_A^2 - m_\alpha^2}{m_A^2 + m_\alpha^2}
\,,
\qquad
E'_A
=
m_A + \frac{\left( m_A - m_\alpha \right)^2}{2 m_\alpha}
\,.
\label{018}
\end{equation}

In conclusion,
we have shown that Lorentz invariance implies that in general
flavor neutrinos in oscillation experiments
are superpositions of massive neutrinos with different momenta and energies.
For each production process (unless $\xi=1/2$),
there is either a Lorentz frame in which
the different massive neutrinos have
equal energy or
a Lorentz frame in which
they have equal momentum.
In the case of neutrinos produced in two-body decays
there is a Lorentz frame in which
the different massive neutrinos have
equal energy.
However,
let us emphasize that such a frame does not have any other attractive property
and is not useful in the calculation
of the flavor transition probability,
because in general it does not correspond to the laboratory frame
and depends on the energy of the decaying particle,
that usually is not monochromatic.

\newpage


\begin{thebibliography}{10}

\bibitem{Pontecorvo-mesonium-57}
B.~Pontecorvo,
\newblock Zh. Eksp. Teor. Fiz. {\bf 33}, 549 (1957),
\newblock [Sov. Phys. JETP \textbf{6}, 429 (1958)].

\bibitem{Pontecorvo-58}
B.~Pontecorvo,
\newblock Zh. Eksp. Teor. Fiz. {\bf 34}, 247 (1958),
\newblock [Sov. Phys. JETP \textbf{7}, 172 (1958)].

\bibitem{Pontecorvo-68}
B.~Pontecorvo,
\newblock Zh. Eksp. Teor. Fiz. {\bf 53}, 1717 (1967),
\newblock [Sov. Phys. JETP \textbf{26}, 984 (1968)].

\bibitem{Gribov-Pontecorvo-69}
V.~N. Gribov and B.~Pontecorvo,
\newblock Phys. Lett. {\bf B28}, 493 (1969).

\bibitem{Eliezer-Swift-MIXING-76}
S.~Eliezer and A.~R. Swift,
\newblock Nucl. Phys. {\bf B105}, 45 (1976).

\bibitem{Fritzsch-Minkowski-OSCILLATIONS-76}
H.~Fritzsch and P.~Minkowski,
\newblock Phys. Lett. {\bf B62}, 72 (1976).

\bibitem{Bilenky-Pontecorvo-AGAIN-76}
S.~M. Bilenky and B.~Pontecorvo,
\newblock Nuovo Cim. Lett. {\bf 17}, 569 (1976).

\bibitem{Bilenky-Pontecorvo-PR-78}
S.~M. Bilenky and B.~Pontecorvo,
\newblock Phys. Rept. {\bf 41}, 225 (1978).

\bibitem{Nussinov-coherence-76}
S.~Nussinov,
\newblock Phys. Lett. {\bf B63}, 201 (1976).

\bibitem{Kayser-oscillations-81}
B.~Kayser,
\newblock Phys. Rev. {\bf D24}, 110 (1981).

\bibitem{Giunti-Kim-Lee-Whendo-91}
C.~Giunti, C.~W. Kim, and U.~W. Lee,
\newblock Phys. Rev. {\bf D44}, 3635 (1991).

\bibitem{Giunti-Kim-Lee-Lee-93}
C.~Giunti, C.~W. Kim, J.~A. Lee, and U.~W. Lee,
\newblock Phys. Rev. {\bf D48}, 4310 (1993), hep-ph/9305276.

\bibitem{Giunti-Kim-QMNO-00}
C.~Giunti and C.~W. Kim,
\newblock Found. Phys. Lett. , in press (2001), hep-ph/0011074.

\bibitem{Giunti-Kim-Lee-Remarks-92}
C.~Giunti, C.~W. Kim, and U.~W. Lee,
\newblock Phys. Rev. {\bf D45}, 2414 (1992).

\bibitem{CWKim-book}
C.~W. Kim and A.~Pevsner,
\newblock {\em Neutrinos in physics and astrophysics} (Harwood Academic Press,
  Chur, Switzerland, 1993),
\newblock {Contemporary Concepts in Physics, Vol. 8}.

\bibitem{BGG-review-98}
S.~M. Bilenky, C.~Giunti, and W.~Grimus,
\newblock Prog. Part. Nucl. Phys. {\bf 43}, 1 (1999), hep-ph/9812360.

\bibitem{Bilenky-Giunti-lep-01}
S.~M. Bilenky and C.~Giunti,
\newblock (2001), hep-ph/0102320.

\bibitem{Winter-81}
R.~G. Winter,
\newblock Lett. Nuovo Cim. {\bf 30}, 101 (1981).

\bibitem{Grossman-Lipkin-spatial-97}
Y.~Grossman and H.~J. Lipkin,
\newblock Phys. Rev. {\bf D55}, 2760 (1997), hep-ph/9607201.

\bibitem{Lipkin-two-slit-00}
H.~J. Lipkin,
\newblock Phys. Lett. {\bf B477}, 195 (2000).

\bibitem{Stodolsky-unnecessary-98}
L.~Stodolsky,
\newblock Phys. Rev. {\bf D58}, 036006 (1998), hep-ph/9802387.

\bibitem{LSND-flight-98}
C. Athanassopoulos \textit{et al.} (LSND Coll.), Phys. Rev. Lett. \textbf{81},
  1774 (1998).

\bibitem{SK-upmu-99}
SuperKamiokande, Y.~Fukuda {\em et~al.},
\newblock Phys. Rev. Lett. {\bf 82}, 2644 (1999), hep-ex/9812014.

\bibitem{Giunti-Kim-Lee-Whendo-98}
C.~Giunti, C.~W. Kim, and U.~W. Lee,
\newblock Phys. Lett. {\bf B421}, 237 (1998), hep-ph/9709494.

\bibitem{Giunti-Kim-Coherence-98}
C.~Giunti and C.~W. Kim,
\newblock Phys. Rev. {\bf D58}, 017301 (1998), hep-ph/9711363.

\end{thebibliography}

\end{document}